\documentclass[aps,superscriptaddress,twocolumn,showpacs,dvips]{revtex4}
\usepackage{feynmp}
\usepackage{amssymb}
\usepackage{epsfig}
\usepackage{graphicx}
\usepackage{subfigure}
\usepackage{hyperref}

\begin{document}

\title{Disorder effects at a nematic quantum critical point in $d$-wave cuprate superconductor}

\author{Jing Wang}
\affiliation{Department of Modern Physics, University of Science and
Technology of China, Hefei, Anhui, 230026, P.R. China}
\author{Guo-Zhu Liu}
\affiliation{Department of Modern Physics, University of Science and
Technology of China, Hefei, Anhui, 230026, P.R. China}
\affiliation{Institut f$\ddot{u}$r Theoretische Physik, Freie
Universit$\ddot{a}$t Berlin, Arnimallee 14, D-14195 Berlin, Germany}
\author{Hagen Kleinert}
\affiliation{Institut f$\ddot{u}$r Theoretische Physik, Freie
Universit$\ddot{a}$t Berlin, Arnimallee 14, D-14195 Berlin, Germany}

\begin{abstract}
A $d$-wave high temperature cuprate superconductor exhibits a
nematic ordering transition at zero temperature. Near the quantum
critical point, the coupling between gapless nodal quasiparticles
and nematic order parameter fluctuation can result in unusual
behaviors, such as extreme anisotropy of fermion velocities. We
study the disorder effects on the nematic quantum critical behavior
and especially on the flow of fermion velocities. The disorders that
couple to nodal quasiparticles are divided into three types: random
mass, random gauge field, and random chemical potential. A
renormalization group analysis shows that random mass and random
gauge field are both irrelevant and thus do not change the fixed
point of extreme velocity anisotropy. However, the marginal
interaction due to random chemical potential destroys this fixed
point and makes the nematic phase transition unstable.
\end{abstract}


\maketitle

%%%%%%%%%%%%%%%%%%%%%%%%%%%%%Main Body%%%%%%%%%%%%%%%%%%%%%%%%%%%%%%%%%%%%%

\section{Introduction}

One important reason that high-temperature cuprate superconductors
are hard to understand is that they have very complicated phase
diagram. The competitions and transitions between different phases
give rise to many unusual properties, and hence have attracted
considerable theoretical and experimental efforts in the past two
decades. Among the widely studied competing orders, the various
phase of the anisotropic electronic liquid are of particular
interests. Kivelson, Fradkin, and Emery proposed that due to the
local electronic phase separation, a number of novel electronic
liquid crystal phases can exist in a doped Mott insulator
\cite{Kivelson}. The simplest of such phases is the electronic
nematic phase, in which the rotational symmetry is broken but the
translational symmetry is preserved. In recent years, the nematic
ordering phase transition has been investigated extensively
\cite{Fradkin, Vojta}. The resistivity anisotropy observed by Ando
$et$ $al.$ in two types of cuprate superconductors provided the
early evidence for the predicted nematic phase \cite{Ando}. More
recently, the neutron-scattering experiments performed in
YBa$_2$Cu$_3$O$_{6.45}$ also pointed to the existence of nematic
phase \cite{Hinkov}. Further evidences came from the observed
in-plane anisotropy of the Nernst effect in the pseudogap region of
YBa$_2$Cu$_3$O$_{y}$ \cite{Daou} and from the scanning tunneling
microscopy experiments performed in the pseudogap region of
Bi$_2$Sr$_2$CaCu$_2$O$_{8+\delta}$ \cite{Lawler}. Interestingly,
there are also compelling experimental indications for the existence
of nematic phase in Sr$_3$Ru$_2$O$_7$ \cite{Mackenzie} and in newly
discovered iron-based superconductor \cite{Chuang}.

In the language of field theory, the nematic phase has an Ising-type
order parameter that can be represented by a real scalar field
$\phi$. However, the dynamics of the system close to the critical
point can not be fully described by an effective $\phi^4$ theory
\cite{Hertz} when there are itinerant electrons. The interaction
between quantum fluctuation of the nematic order parameter and the
itinerant electrons has to be treated carefully. Close to the
quantum critical point, this interaction becomes singular, which was
shown to be able to produce highly unusual, non-Fermi liquid like,
behaviors \cite{Rech, Sunkai, Metlitski, Garst}. The nematic physics
is intimately related to Pomeranchuk instability and has also been
investigated from this point of view \cite{Oganesyan, Metzner}.

Besides the pseudogap phase of underdoped cuprates, it is also
interesting to study the nematic transition that occurs in the
$d$-wave superconducting phase \cite{VojtaSachdev, Kim, Huh, Xu,
Fritz}. This is a new example of quantum phase transitions happening
in the superconducting dome \cite{Castellani, Sachdev09, Zaanen},
which is a widely studied topic. In the superconducting phase, the
nematic order parameter interacts strongly with the gapless nodal
quasiparticles, which are the low-energy excitations of a $d$-wave
superconductor. This interaction remarkably affects the dynamics of
both nodal quasiparticles and nematic order parameter. An early work
of Vojta $et$ $al.$ \cite{VojtaSachdev} presented a detailed
renormalization group (RG) analysis of various types of Yukawa
couplings in the $d$-wave superconducting phase, including nematic
type coupling. More recently, Kim $et$ $al.$ studied the effect of
quantum fluctuations of nematic order parameter on the spectral
properties of nodal quasiparticles \cite{Kim}.

In actual $d$-wave cuprate superconductor, the gapless nodal
quasiparticles have a Fermi velocity $v_F$ and a gap velocity
$v_{\Delta}$, which are not equal. Indeed, the ratio
$v_{\Delta}/v_{F}$ may be as small as $1/20$ \cite{Millis}. This
small ratio plays an important role because it appears in a number
of observable quantities \cite{Millis}. For instance, it was found
\cite{Durst} that the dc thermal conductivity contains the large
inverse of this ratio as
\begin{equation}
\frac{\kappa}{T} \propto \frac{k_{B}^{2}}{\hbar}
\left(\frac{v_{\Delta}}{v_{F}} + \frac{v_{F}}{v_{\Delta}}\right)
\end{equation}
at nearly zero temperature. This easily accessible material property
is universal --- it is independent of the amount of disorder
\cite{Durst}. This universality was confirmed by transport
measurements \cite{Taillefer}. Since the inverse of
$v_{\Delta}/v_{F}$ is so large it completely dominates
${\kappa}/{T}$.

An interesting problem is how the velocity ratio is influenced by
the nematic phase transition. Recently, Huh and Sachdev \cite{Huh}
studied this problem by making a careful RG analysis within an
effective field theory of nematic ordering transition. They found
that $v_{\Delta}/v_{F}$ flows to a fixed point with
$v_{\Delta}/v_{F} \rightarrow 0$, i.e., the inverse velocity ratio
$v_{F}/v_{\Delta}$ diverges. Therefore, the nematic ordering
transition in $d$-wave superconductor is accompanied by the
appearance of an extreme velocity anisotropy. Since the diverging
velocity ratio $v_F/v_{\Delta}$ enters various physical properties,
the predicted extreme anisotropy should have observable effects. In
particular, the low-temperature dc thermal conductivity is expected
to be significantly enhanced near the critical point. By using a
Boltzmann equation approach, Fritz and Sachdev \cite{Fritz}
calculated the thermal conductivity enhancement near the nematic
quantum critical point due to the divergence of $v_{F}/v_{\Delta}$.
If this enhancement were observed in transport experiments at
certain doping concentration, this would serve as an important
evidence for the existence of nematic transition in $d$-wave cuprate
superconductor.

When studying the low-temperature transport properties of an
interacting electron system, it is hardly possible to ignore the
disorder effects. First of all, the fermions are always scattered by
certain amount of disorder in any realistic physical system.
Moreover, although the elastic scattering due to quenched disorder
is less important at high temperature, it dominates over the
inelastic scattering due to inter-particle interactions at very low
temperature. The disorder effects should be taken into account when
calculating the low-temperature thermal conductivity. The nematic
order parameter fluctuation can lead to significant enhancement of
thermal conductivity only when the fixed point of extreme velocity
anisotropy is stable against disorder scattering. If the fixed point
is changed or even destroyed by disorder, the thermal conductivity
enhancement will not occur in practice. It is therefore crucial to
examine the disorder effects on the RG flow of fermion velocities,
especially on the stability of extreme anisotropy of velocities.

In general, the disorders coupled to gapless nodal quasiparticles in
$d$-wave superconductor can be divided into three types: random
chemical potential, random gauge field, and random mass. The
difference comes from the different Pauli matrices used to define
the fermion-disorder interacting terms. The effects of these
disorders on the low-temperature transport properties of nodal
quasiparticles have been discussed extensively \cite{Nersesyan,
Altland}. These disorders will alter the RG flows of fermion
velocities. On the other hand, the RG flow of strength parameters of
fermion-disorder couplings are determined by fermion velocities, and
thus should be calculated self-consistently with the flow of fermion
velocities.

In this paper, we present a RG analysis of the interplay between
nematic order parameter fluctuation and disorder scattering. We
derive a series of coupled RG equations of fermion Fermi velocity
$v_F$, gap velocity $v_\Delta$, and disorder strength parameter $g$.
In the cases of random mass and random gauge field, the
corresponding disorder strength parameters both flow to zero at low
energy. Therefore, these two kinds of disorders do not change the
flow of fermion velocities and hence the fixed point of extreme
velocity anisotropy is stable. However, the strength parameter of
random chemical potential remains a constant even down to the lowest
energy, and thus is able to modify the fermion velocities
significantly. We found that the fermion velocities do not flow to
any fixed points, but indeed oscillate rapidly between positive and
unphysical negative values. This implies that the extreme anisotropy
fixed point is destroyed and the nematic phase transition may become
unstable due to random chemical potential.

In Sec. \ref{Model}, we define the model action in the presence of
both nematic order parameter fluctuation and disorder. In Sec.
\ref{part_3}, we calculate the fermion self-energy corrections due
to nematic order parameter and disorder scattering. The
fermion-disorder vertex corrections due to nematic and disorder
interactions are also computed in this section. In Sec.
\ref{RG_equations}, we make the RG analysis and obtain the
self-consistent RG equations for fermion velocities and disorder
strength parameter. These equations are solved both analytically and
numerically in Sec. \ref{Stability_discuss}. From the solutions, we found that the fixed
point of extreme velocity anisotropy is not changed by random mass
and random gauge field. However, the random chemical potential
destroys this fixed point and indeed makes the nematic phase
transition unstable. In Sec. \ref{summary}, we briefly summarize the results
obtained in this paper and discuss the possible experimental
detection of the predicted extreme velocity anisotropy.

\section{Model}\label{Model}

We start from the following action
\begin{eqnarray}
S &=& S_{\psi} + S_{\phi} + S_{\psi\phi},
\end{eqnarray}
where the free action for nodal quasiparticles is
\begin{eqnarray}\!\!\!\!\!\!\!\!\!\!\!\!\!\!\!\!\!\!\!
S_{\psi} \!&=&\!\!\! \int\frac{d^{2}\mathbf{k}}{(2\pi)^{2}}\frac{d\omega}{2\pi}
\psi^{\dagger}_{1a}(-i\omega+v_{F}k_{x}\tau^{z} +
v_{\Delta}k_{y}\tau^x)\psi_{1a} \nonumber \\
&&\hspace{-1.4em} +\int\frac{d^{2}\mathbf{k}}{(2\pi)^2}\frac{d\omega}{2\pi}
\psi^{\dagger}_{2a}(-i\omega+v_{F}k_{y}\tau^{z} +
v_{\Delta}k_{x}\tau^{x})\psi_{2a},
\end{eqnarray}
where $\tau^{(x,y,z)}$ denote Pauli matrices. The linear dispersion
of Dirac fermions originates from the $d_{x^2 - y^2}$-wave symmetry
of the energy gap of cuprate superconductor. Here, the spinor
$\psi^{\dagger}_{1}$ represents nodal quasiparticles excited from
the $(\frac{\pi}{2},\frac{\pi}{2})$ and
$(-\frac{\pi}{2},-\frac{\pi}{2})$ nodal points, and
$\psi^{\dagger}_{2}$ the other two nodal points \cite{VojtaSachdev}.
The repeated spin index $a$ is summed from 1 to $N_f$, the number of
fermion spin components. The ratio $v_{\Delta}/v_{F} \approx 1/20$
between Fermi velocity and gap velocity is determined by experiments
\cite{Millis}. The effective action $S_{\phi}$ describes the Ising
type nematic order parameter, which is expanded (for notational
simplicity) in real space as
\begin{eqnarray}
S_{\phi} = \int d^2\mathbf{x}d\tau\Big\{\frac{1}{2}(\partial\tau
\phi)^2 + \frac{c^2}{2}(\nabla\phi)^2 +
\frac{r}{2}\phi^2+\frac{u_0}{24}\phi^4\Big\},
\end{eqnarray}
where $\tau$ is imaginary time and $c$ is velocity. The mass
parameter $r$ tunes the nematic phase transition with $r=0$ defining
the quantum critical point. The parameter $u_0$ is the quartic
self-interaction strength. The nematic order parameter couples to
nodal quasiparticles via the Yukawa term
\begin{eqnarray}
S_{\psi\phi} = \int d^2\mathbf{x}d\tau\{\lambda_0
\phi(\psi^{\dagger}_{1a}\tau^{x}\psi_{1a} +
\psi^{\dagger}_{2a}\tau^{x}\psi_{2a})\}.
\end{eqnarray}

\begin{figure}
\includegraphics[width=1.99in]{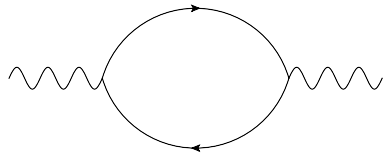}
\caption{The polarization function for nematic order parameter. The
solid line represents the fermion propagator and wavy line
represents the boson propagator.}\label{nem_polarization}
\end{figure}

Following Huh and Sachdev \cite{Huh}, we now perform the RG analysis
in the framework of a $1/N_f$ expansion. The inverse of the free
propagator of the nematic order parameter field behaves as $q^2 +
r$. After taking into account the polarization effects, there will
be an additional linear $q$-term. At low energy regime, the $q$-term
dominates over the $q^2$-term, which then can be neglected. Near the
quantum critical point, we keep only the mass term and assume that
$\phi \longrightarrow \phi/\lambda_0$ and $r\longrightarrow N_f r
\lambda^2_0$, leading to
\begin{eqnarray}
S = S_{\psi} \!+\!\! \int\! \!d^2\mathbf{x} d \tau \Big\{\frac{N_fr}{2}\phi^2 \!+
\phi[\psi^{\dagger}_{1a} \tau^x \psi_{1a}\! +
\psi^{\dagger}_{2a}\tau^x \psi_{2a}]\Big\}\!.
\end{eqnarray}
After integrating out fermion degrees of freedom, the effective
action for the scalar field becomes
\begin{eqnarray}
\frac{S_{\phi}}{N} = \frac{1}{2}\int \frac{d^3q}{(2\pi)^3}[r+\Pi(q)]|\phi(q)|^{2} +
\mathcal{O}(\phi^{4}).
\end{eqnarray}
The lowest-order Feynman diagram for the polarization function is
shown in Fig. \ref{nem_polarization} and symbolizes the integral
\begin{eqnarray}
\Pi(\mathbf{q},\epsilon) =
\int\frac{d^{2}\mathbf{k}}{(2\pi)^{2}}\frac{d\omega}{2\pi}
\mathrm{Tr}[\tau^{x}G^{0}_{\psi}(\mathbf{k},\omega)\tau^{x}
G^{0}_{\psi}(\mathbf{k+q},\omega+\epsilon)], \nonumber
\end{eqnarray}
where the free fermion propagator is
\begin{eqnarray}
G^{0}_{\psi}(\mathbf{k},\omega) = \frac{1}{-i\omega +
v_{F}k_{x}\tau^{z} + v_{\Delta}k_{y}\tau^{x}}.
\end{eqnarray}
As shown previously \cite{Huh}, the propagator for the nematic order
parameter is given by
\begin{eqnarray}
G_{\phi}^{-1}(\mathbf{q},\epsilon) &=&
\Pi(\mathbf{q},\epsilon) \nonumber \\
&=& \frac{1}{16v_{F}v_{\Delta}}
\frac{(\epsilon^{2}+v_{F}^{2}q_{x}^{2})}{(\epsilon^2 +
v_{F}^{2}q_{x}^{2}+v_{\Delta}^{2}q_{y}^{2})^{1/2}} \nonumber \\
&& +\frac{1}{16v_{F}v_{\Delta}}\frac{(\epsilon^{2}+v_{F}^{2}
q_{y}^{2})}{(\epsilon^{2}+v_{F}^{2}q_{y}^{2} +
v_{\Delta}^{2}q_{x}^{2})^{1/2}}
\end{eqnarray}
in the vicinity of nematic quantum critical point $r=0$.

Disorders are present in almost all realistic condensed matter
systems and play important roles in determining the low-temperature
behaviors. In the present problem, the nodal quasiparticles can
interact with three types of random potentials, which represent
different disorder scattering processes. According to the coupling
between nodal quasiparticle and disorders, there are three types of
random fields in $d$-wave superconductors: random mass, random
chemical potential, and random gauge potential. All these types of
disorders have been investigated in the contexts of $d$-wave cuprate
superconductor \cite{Nersesyan, Altland}, quantum Hall effect
\cite{Ludwig}, and graphene \cite{Stauber, Aleiner}. In the general
analysis to follow, we shall consider the three types of disorders.

\begin{figure}
\includegraphics[width=3.4in]{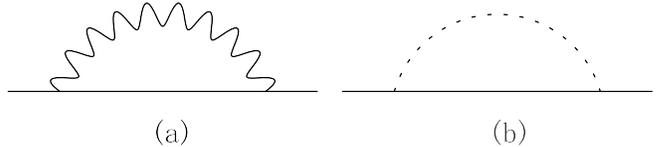}
\caption{One loop fermion self-energy correction due to (a) nematic
order parameter fluctuation and (b) disorder. The dashed line
represents disorder scattering.}\label{nem_dis_self}
\end{figure}

The fermion field couples to a random field $A(\mathbf{x})$
as
\begin{eqnarray}
\int d^2 \mathbf{x}
\psi^{\dagger}(\mathbf{x})\Gamma\psi(\mathbf{x})A(\mathbf{x}),
\end{eqnarray}
For random chemical potential, the matrix $\Gamma$ is $\Gamma =
\mathrm{I}$. For a random mass it is, $\Gamma = \tau^{y}$, and for a
random gauge field $\Gamma = (\tau^{x},\tau^{z})$. The random
potential $A(\mathbf{x})$ is assumed to be a quenched, Gaussian
white noise field with the correlation functions
\begin{eqnarray}
\langle A(\mathbf{x})\rangle = 0; \hspace{0.5cm} \langle
A(\mathbf{x}_1)A(\mathbf{x}_2)\rangle = g
v_{\Gamma}^{2}\delta^2(\mathbf{x}_1 - \mathbf{x}_2).
\end{eqnarray}
The dimensionless parameter $g$ represents the concentration of
impurity, and the parameter $v_\Gamma$ measures the strength of a
single impurity. It will be convenient to redefine the random
potential as $A(\mathbf{x})\rightarrow v_{\Gamma}A(\mathbf{x})$,
and then write the fermion-disorder interaction term as
\cite{Stauber}
\begin{eqnarray}
S_{\mathrm{dis}} = v_{\Gamma}\int d^2 \mathbf{x}
\psi^{\dagger}(\mathbf{x})\Gamma\psi(\mathbf{x})A(\mathbf{x}),
\end{eqnarray}
with the random potential distribution
\begin{eqnarray}\label{A_equation}
\langle A(\mathbf{x})\rangle = 0; \hspace{0.5cm} \langle
A(\mathbf{x}_1)A(\mathbf{x}_2)\rangle = g\delta^2(\mathbf{x}_1 -
\mathbf{x}_2).
\end{eqnarray}
Now the RG flow of disorder strength can be calculated by studying
the vertex correction to the fermion-disorder interaction term.
After a Fourier transformation, the corresponding action has the
form
\begin{eqnarray}
S_{\mathrm{dis}} = v_{\Gamma}\int d^{2}\mathbf{k}
d^{2}\mathbf{k}_1d\omega \psi^{\dagger}(\mathbf{k},\omega)\Gamma
\psi(\mathbf{k}_1,\omega)A(\mathbf{k-k_1}).
\end{eqnarray}
This action will be analyzed together with the actions (3), (6), and
(7). In order to perform perturbative expansion, we assume that $g$
and $v_\Gamma$ are both small in magnitude, corresponding to the
weak disorder case.

\section{Fermion self-energy and fermion-disorder vertex corrections}\label{part_3}

According to the Dyson equation, the interactions induce a
self-energy correction to the free propagator of Dirac fermion,
yielding
\begin{eqnarray}
G^{-1}_{\psi}(\mathbf{k},\omega) &=& -i\omega + v_F k_x\tau^z +
v_{\Delta} k_{y}\tau^{x} \nonumber \\
&& -\Sigma_{\mathrm{nm}}(\mathbf{k},\omega) -
\Sigma_{\mathrm{dis}}(\mathbf{k},\omega),
\end{eqnarray}
where self-energy functions $\Sigma_{\mathrm{nm}}$ and
$\Sigma_{\mathrm{dis}}$ come from nematic ordering and disorder
scattering, respectively. To the leading order, the corresponding
Feynman diagrams are presented in Fig. \ref{nem_dis_self}.

The nematic self-energy $\Sigma_{\mathrm{nm}}$ has already been
obtained by Huh and Sachdev \cite{Huh}, who found that
\begin{eqnarray}
\frac{d\Sigma_{\mathrm{nm}}(\mathbf{k},\omega)}{d\ln\Lambda} =
C_1(-i\omega) + C_2 v_F k_x \tau^z + C_3 v_{\Delta} k_y \tau^x,
\end{eqnarray}
where
\begin{eqnarray}
C_1 &=& \frac{2(v_\Delta/v_F)}{N_f \pi^3}\int^{\infty}_{-\infty}dx
\int^{2\pi}_{0} d \theta \nonumber \\
&&\times\frac{x^2-\cos^2\theta-(v_\Delta/v_F)^2
\sin^2\theta}{(x^2+\cos^2\theta+(v_\Delta/v_F)^2
\sin^2\theta)^2}\mathcal {G}(x,\theta), \\
C_2 &=& \frac{2(v_\Delta/v_F)}{N_f \pi^3}\int^{\infty}_{-\infty}dx
\int^{2\pi}_{0} d \theta  \nonumber \\
&&\times\frac{\cos^2\theta-x^2-(v_\Delta/v_F)^2
\sin^2\theta}{(x^2+\cos^2\theta+(v_\Delta/v_F)^2 \sin^2\theta)^2}\mathcal
{G}(x,\theta), \\
C_3 &=& \frac{2(v_\Delta/v_F)}{N_f \pi^3}\int^{\infty}_{-\infty}dx
\int^{2\pi}_{0} d \theta  \nonumber \\
&&\times\frac{x^2+\cos^2\theta-(v_\Delta/v_F)^2
\sin^2\theta}{(x^2+\cos^2\theta+(v_\Delta/v_F)^2 \sin^2\theta)^2}\mathcal
{G}(x,\theta), \\
\mathcal{G}^{-1} &=& \frac{x^2+\cos^2\theta}{\sqrt
{x^2+\cos^2\theta+(v_\Delta/v_F)^2 \sin^2\theta}} \nonumber \\
&& + \frac{x^2+\sin^2\theta}{\sqrt{x^2 +
\sin^2\theta+(v_\Delta/v_F)^2\cos^2\theta}}.
\end{eqnarray}
The computational details of $C_{1,2,3}$ are presented in the
appendix.

The fermion self-energy due to disorder
$\Sigma_{\mathrm{dis}}(i\omega)$ can be computed as
\begin{eqnarray}
\Sigma_{\mathrm{dis}}(i\omega) &=& gv^2_\Gamma
\int\frac{d^{2}\mathbf{k}}{(2\pi)^2}\Gamma
G^{0}_{\psi}(\mathbf{k},\omega)\Gamma \nonumber \\
&=& \frac{g v_{\Gamma}^{2}}{2\pi v_{F}v_{\Delta}}i\omega\ln\Lambda.
\end{eqnarray}
From this expression, we know that $\Sigma_{\mathrm{dis}}(i\omega)$
has the same result for all possible expressions of $\Gamma$.
Another important feature is that $\Sigma_{\mathrm{dis}}(i\omega)$
is independent of momentum, which reflects the fact that the
quenched disorder is static. It is easy to have
\begin{eqnarray}
\frac{d\Sigma_{\mathrm{dis}}(i\omega)}{d\ln\Lambda} = C_g i\omega,
\end{eqnarray}
where
\begin{eqnarray}
C_g = \frac{g v_{\Gamma}^{2}}{2\pi v_{F} v_{\Delta}}.
\end{eqnarray}

\begin{figure}
\includegraphics[width=3.35in]{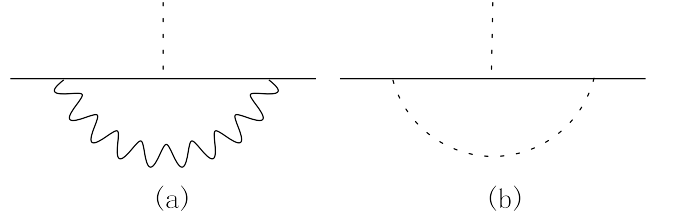}
\caption{Fermion-disorder vertex correction due to (a) nematic order
parameter and (b) disorder
parameter.}\label{nem_dis_vertex}
\end{figure}

The fermion-disorder interaction parameter $v_\Gamma$ is also
subjected to RG flow. To get its flow equation, we need to calculate
the fermion-disorder vertex corrections. Formally, the vertex
correction has the form
\begin{equation}
v_{\Gamma}\Gamma' = v_{\Gamma}\Gamma + V_{\mathrm{nm}} +
V_{\mathrm{dis}},
\end{equation}
where $V_{\mathrm{nm}}$ represents the vertex correction due to
nematic order parameter fluctuation and $V_{\mathrm{dis}}$
represents the vertex correction due to disorder interaction. The
corresponding diagrams are shown in Fig. \ref{nem_dis_vertex}. They
will be calculated explicitly in the following for all three kinds
of disorders.

\subsubsection{Random chemical potential}

We first calculate the vertex correction due to nematic ordering. To
this end, we employ the method proposed by Huh and Sachdev
\cite{Huh}. At zero external momenta and frequencies, the vertex
correction is expressed as
\begin{eqnarray}
V_{\mathrm{nm}} = v_{\Gamma} \int\frac{d^{3}Q}{(2\pi)^3}H(Q)
\mathcal{K}^{3}(\frac{\mathbf{q}^{2}}{\Lambda^{2}}).
\end{eqnarray}
There is an useful formula \cite{Huh},
\begin{eqnarray}
\frac{dV_{\mathrm{nm}}}{d\ln\Lambda} = v_{\Gamma}
\frac{v_{F}}{8\pi^3}\int^{\infty}_{-\infty}dx \int ^{2\pi}_{0}
d\theta H(\hat{Q}),
\end{eqnarray}
where
\begin{eqnarray}
H(\hat{Q}) &=& \frac{1}{N_f}\tau^x\frac{1}{(-iv_{F}x +
v_{F}\cos\theta\tau^{z} + v_{\Delta}\sin\theta\tau^x)}\mathrm{I}
\nonumber \\
&& \times \frac{1}{(-iv_{F}x + v_{F}\cos\theta\tau^{z} +
v_{\Delta}\sin\theta\tau^{x})}\tau^{x}\frac{1}{\Pi(\hat{Q})}.
\nonumber \\
\end{eqnarray}
Here, the matrix $\mathrm{I}$ corresponds to the coupling between
Dirac fermion and random chemical potential. It will be replaced by
$\tau^{y}$ in the case of random mass and $\tau^{x,z}$ in the case
of random gauge field. After straightforward computation, we have
\begin{eqnarray}
\frac{dV_{\mathrm{nm}}}{d\ln\Lambda} = C_{5}v_{\Gamma}\mathrm{I},
\end{eqnarray}
where
\begin{eqnarray}
C_{5} &=& -\frac{2(v_\Delta/v_F)}{N_f\pi^3}\int^\infty_{-\infty} dx
\int^{2\pi}_{0} d\theta \nonumber \\
&& \times \frac{(x^2 - \cos^{2}\theta -
(v_{\Delta}/v_{F})^{2}\sin^{2}\theta)} {(x^{2} + \cos^{2}\theta +
(v_{\Delta}/v_{F})^{2}\sin^{2}\theta)^{2}}\mathcal{G}(x,\theta)
\nonumber \\
&=& -C_1.
\end{eqnarray}

The vertex correction due to averaging over disorder is
\begin{eqnarray}
V_{\mathrm{dis}} = gv_{\Gamma}^{2}
\int\frac{d^{2}\mathbf{p}}{(2\pi)^{2}} \mathrm{I}
G^0_{\psi}(\omega,\mathbf{p})v_{\Gamma}\mathrm{I}
G^0_{\psi}(\omega,\mathbf{p+k})\mathrm{I}.
\end{eqnarray}
Again, the matrix $\mathrm{I}$ should be replaced by certain Pauli
matrix in the case of random mass or random gauge field. Taking the
external momentum $\mathbf{k}=0$ and keeping only the leading
divergent term, we have
\begin{eqnarray}
\frac{dV_{\mathrm{dis}}}{d\ln\Lambda} =
C_{\Gamma}v_{\Gamma}\mathrm{I},
\end{eqnarray}
where
\begin{eqnarray}
C_{\Gamma} = \frac{v_{\Gamma}^{2} g}{2\pi v_{F}v_{\Delta}} = C_{g}.
\end{eqnarray}

\subsubsection{Random mass}

The calculation of vertex correction in the case of random mass
parallels the process presented above, so we just state the final
result. The nematic ordering induced vertex correction is
\begin{eqnarray}
\frac{dV_{\mathrm{nm}}}{d\ln\Lambda} = C_{6}v_{\Gamma}\tau^{y},
\end{eqnarray}
where
\begin{eqnarray}
C_{6} &=& \frac{2(v_\Delta/v_F)}{N_f\pi^3}\int^\infty_{-\infty} dx
\int ^{2\pi}_0d \theta \nonumber \\
&& \times \frac{(x^2+\cos^2\theta +
(v_{\Delta}/v_{F})^2\sin^2\theta)}{(x^2+\cos^2\theta +
(v_{\Delta}/v_{F})^2\sin^2\theta)^2}\mathcal{G}(x,\theta) \nonumber \\
&=& C_{3}-C_{1}-C_{2}.
\end{eqnarray}
The disorder induced vertex correction is
\begin{eqnarray}
\frac{dV_{\mathrm{dis}}}{d\ln\Lambda} =
-C_{\Gamma}(v_{\Gamma}\tau^{y}),
\end{eqnarray}
where
\begin{eqnarray}
C_{\Gamma} = \frac{v_{\Gamma}^{2} g}{2\pi v_{F}v_{\Delta}} = C_{g}.
\end{eqnarray}

\subsubsection{Random gauge potential}

The random gauge potential has two components, characterized by
$\tau^x$ and $\tau^z$ respectively. For the $\tau^x$ component, the
nematic ordering contribution to vertex correction is
\begin{eqnarray}
\frac{dV_{\mathrm{nm}}}{d\ln\Lambda} = C_{4A}v_{\Gamma}\tau^{x},
\end{eqnarray}
where
\begin{eqnarray}
C_{4A} &=& -\frac{2(v_{\Delta}/v_{F})}{N_f\pi^3}
\int^\infty_{-\infty} dx \int^{2\pi}_{0}d\theta \nonumber \\
&& \times \frac{(x^{2}+\cos^{2}\theta -
(v_{\Delta}/v_{F})^{2}\sin^{2}\theta)} {(x^{2} + \cos^{2}\theta +
(v_{\Delta}/v_{F})^{2}\sin^{2}\theta)^{2}} \mathcal{G}(x,\theta)
\nonumber \\
&=& -C_{3}.
\end{eqnarray}
For the $\tau_z$ component, we have
\begin{eqnarray}
\frac{dV_{\mathrm{nm}}}{d\ln\Lambda} = C_{4B}v_{\Gamma}\tau^{z},
\end{eqnarray}
where
\begin{eqnarray}
C_{4B} &=& -\frac{2(v_{\Delta}/v_{F})}{N_{f}\pi^{3}}
\int^{\infty}_{-\infty}dx\int^{2\pi}_{0}d\theta \nonumber \\
&& \times \frac{(x^{2} - \cos^{2}\theta +
(v_{\Delta}/v_{F})^{2}\sin^{2}\theta)} {(x^{2} + \cos^{2}\theta +
(v_{\Delta}/v_{F})^{2}\sin^{2}\theta)^{2}} \mathcal{G}(x,\theta)
\nonumber \\
&=& -C_{2}.
\end{eqnarray}

The disorder contribution can be calculated similarly. To both the
$\tau^x$ and $\tau^z$ component, we have
\begin{eqnarray}
V_{\mathrm{dis}}(\omega) = \mathrm{finite},
\end{eqnarray}
so that
\begin{eqnarray}
\frac{dV_{\mathrm{dis}}}{d\ln\Lambda} = 0.
\end{eqnarray}
for both two components.

\section{RG equations for fermion velocities and disorder strength}\label{RG_equations}

In order to perform the RG analysis of the fermion velocity and
disorder strength, it is convenient to make the following scaling
transformations \cite{Huh, Shankar}
\begin{eqnarray}
k_i &=& k'_ie^{-l}, \\
\omega &=& \omega'e^{-l}, \\
\psi_{1,2}(\mathbf{k},\omega) &=& \psi'_{1,2}(\mathbf{k'},\omega')e^{\frac{1}{2}
\int^{l}_{0}(4 - \eta_{f})dl}, \\
\phi(\mathbf{q},\epsilon) &=&
\phi'(\mathbf{q'},\epsilon')e^{\frac{1}{2}\int^{l}_{0} (5 -
\eta_{b})dl},
\end{eqnarray}
where $i=x,y$ and $b = e^{-l}$ with $l>0$. The parameters $\eta_f$
and $\eta_b$ will be determined by the self-energy and
nematic-fermion vertex corrections. Note the energy is required to
scale in the same way as the momentum, so the fermion velocities are
forced to flow under RG transformations.

In the spirit of RG theory \cite{Shankar}, to specify how a field
operator transforms when the energy and momenta are re-scaled, the
standard method is to require that its kinetic term remains
invariant. In the present problem, however, the random potential
$A(\mathbf{x})$ does not have an own kinetic term. In order to find
out its scaling behavior, we write the Gaussian white noise
distribution in the momentum space as
\begin{eqnarray}\label{Gauss_distribution}
\langle A(\mathbf{k}_1)A(\mathbf{k}_2) \rangle =
g\delta^2(\mathbf{k}_1 + \mathbf{k}_2).
\end{eqnarray}
When the momentum $\mathbf{k}$ becomes $b \mathbf{k}$, the delta
function is rescaled to
\begin{eqnarray}
\delta^2(\mathbf{k}_1 + \mathbf{k}_2)\rightarrow
\delta^2(b\mathbf{k}_1 + b\mathbf{k}_2) = b^{-2}
\delta^2(\mathbf{k}_1 + \mathbf{k}_2).
\end{eqnarray}
If we require that the disorder distribution Eq.
(\ref{Gauss_distribution}) is invariant under scaling
transformations, then the random potential should transform as
\begin{eqnarray}
A(\mathbf{k}) \rightarrow b^{-1} A(\mathbf{k}).
\end{eqnarray}
Now we have to assume that
\begin{eqnarray}
A(\mathbf{k}) = A'(\mathbf{k'})e^{l}.
\end{eqnarray}

According to the RG technique presented in \cite{Shankar}, the
momentum shell between $b\Lambda$ and $\Lambda$ will be integrated
out, while keeping the $-i\omega$ term invariant. From the nematic
ordering and disorder contributions to fermion self-energy function,
we have
\begin{eqnarray}
&& \int^{b\Lambda}d^{2}\mathbf{k} d\omega \psi^{\dagger}[-i\omega -
C_1(-i\omega)\ln\frac{\Lambda}{b\Lambda} +
C_g(-i\omega)\ln\frac{\Lambda}{b\Lambda}]\psi \nonumber \\
&& = \int^{b\Lambda}d^{2}\mathbf{k} d\omega \psi^{\dagger}
(-i\omega)[1+(C_g - C_1)l]\psi \nonumber \\
&& \approx \int^{b\Lambda}d^{2}\mathbf{k} d\omega
\psi^{\dagger}(-i\omega)e^{(C_g - C_1)l}\psi.
\end{eqnarray}
After the scaling transformation, this term should go back to the
free form, so that
\begin{eqnarray}
\eta_{f} = C_{g} - C_{1}.
\end{eqnarray}
The kinetic terms should also be kept invariant under scaling
transformation, which leads to
\begin{eqnarray}
\frac{dv_{F}}{dl} &=& (C_{1}-C_{2}-C_{g})v_{F}, \\
\frac{dv_{\Delta}}{dl} &=& (C_{1}-C_{3}-C_{g})v_{\Delta}.
\end{eqnarray}
Based on these expressions, the ratio between gap velocity and Fermi
velocity is
\begin{eqnarray}
\frac{d(v_{\Delta}/v_{F})}{dl} = (C_{2}-C_{3})(v_{\Delta}/v_{F}).
\end{eqnarray}

The disorder strength $g$ appears in the above expressions. Due to
the interplay of nematic ordering and disorder, this parameter also
flows under RG transformation. The flow equation depends on the type
of disorder, which will be studied in the following.

We first consider the case of random chemical potential. The bare
fermion-disorder action is
\begin{eqnarray}
v_{\Gamma}\int d^{2}\mathbf{k}  d^{2}\mathbf{k}_1d\omega
\psi^{\dagger}(\mathbf{k},\omega)\Gamma
\psi(\mathbf{k}_1,\omega)A(\mathbf{k-k_1}).
\end{eqnarray}
Including corrections due to nematic and disorder interactions
yields
\begin{eqnarray}
&&\int^{b\Lambda} d^{2}\mathbf{k}
d^{2}\mathbf{k}_1d\omega\psi^{\dagger}(\mathbf{k},\omega)[v_{\Gamma}\mathrm{I}
- C_{1}v_{\Gamma}\mathrm{I}\ln\frac{\Lambda}{b\Lambda}\nonumber \\
&& + C_{g}v_{\Gamma}\mathrm{I}\ln\frac{\Lambda}{b\Lambda}]
\psi(\mathbf{k}_1,\omega)A(\mathbf{k-k_1}) \nonumber \\
&=& \int^{b\Lambda}d^{2}\mathbf{k}d^{2}\mathbf{k}_1d\omega
\psi^{\dagger}(\mathbf{k},\omega)v_{\Gamma}\mathrm{I}[1+(C_{g}-C_{1})l]
\nonumber \\
&& \times \psi(\mathbf{k}_1,\omega) A(\mathbf{k-k_1}) \nonumber \\
&\approx& \int^{b\Lambda}d^{2}\mathbf{k}d^{2}\mathbf{k}_1d\omega
\psi^{\dagger}(\mathbf{k},\omega)v_{\Gamma}\mathrm{I}e^{(C_{g}-C_{1})l}
\nonumber \\
&& \times \psi(\mathbf{k}_1,\omega) A(\mathbf{k-k_1}).
\end{eqnarray}
After redefining energy, momentum, and field operators, we have
\begin{eqnarray}
&& \int^{\Lambda}d^{2}\mathbf{k'}d^{2}\mathbf{k'}_1 d\omega'
\psi'^{\dagger}(\mathbf{k'},\omega')v_{\Gamma}\mathrm{I} \nonumber \\
&& \times e^{(C_{g}-C_{1})l} \psi'(\mathbf{k'}_1,\omega')
e^{-\eta_{f}l} A'(\mathbf{k}'-\mathbf{k}'_1).
\end{eqnarray}
Since $\eta_{f} = C_{g} - C_{1}$, it is easy to obtain the following
RG flow equation for $v_{\Gamma}$,
\begin{eqnarray}
\frac{dv_{\Gamma}}{dl}=0.
\end{eqnarray}
Apparently, the parameter $v_{\Gamma}$ does not flow and thus can be
simply taken to be a constant.

In the case of random mass, the flow equations for fermion
velocities have the same expressions as Eq.(53) and Eq.(54).
However, the flow equation for disorder strength is different from
Eq.(58), and has the form
\begin{eqnarray}
\frac{dv_{\Gamma}}{dl} = (C_{3} - C_{2} - 2C_{g})v_{\Gamma},
\end{eqnarray}
which couples self-consistently to flow equations of fermion
velocities.

Following the steps presented above, we find the following RG
equations in the case of random gauge potential
\begin{eqnarray}
\frac{d v_F}{dl}&=&(C_1 - C_2 - C_{gi})v_F,\\
\frac{d v_\Delta}{dl}&=&(C_1 - C_3 - C_{gi})v_\Delta,
\end{eqnarray}
which couple to the flow equations of disorder strength
\begin{eqnarray}
\frac{dv_{\Gamma1}}{dl}&=&[(C_1-C_{g1})-C_3]v_{\Gamma1}, \\
\frac{dv_{\Gamma2}}{dl}&=&[(C_1-C_{g2})-C_2]v_{\Gamma2},
\end{eqnarray}
where
\begin{eqnarray}
C_{g_i} = \frac{v_{\Gamma i}^2 g}{2\pi v_F
v_\Delta},\hspace{0.2cm}i=1,2.
\end{eqnarray}
Here, the equations denoted by $i=1,2$ correspond to the $\tau^x$
and $\tau^z$ components of random gauge potential, respectively.

\begin{figure}[t]
  \centering
       \epsfig{file=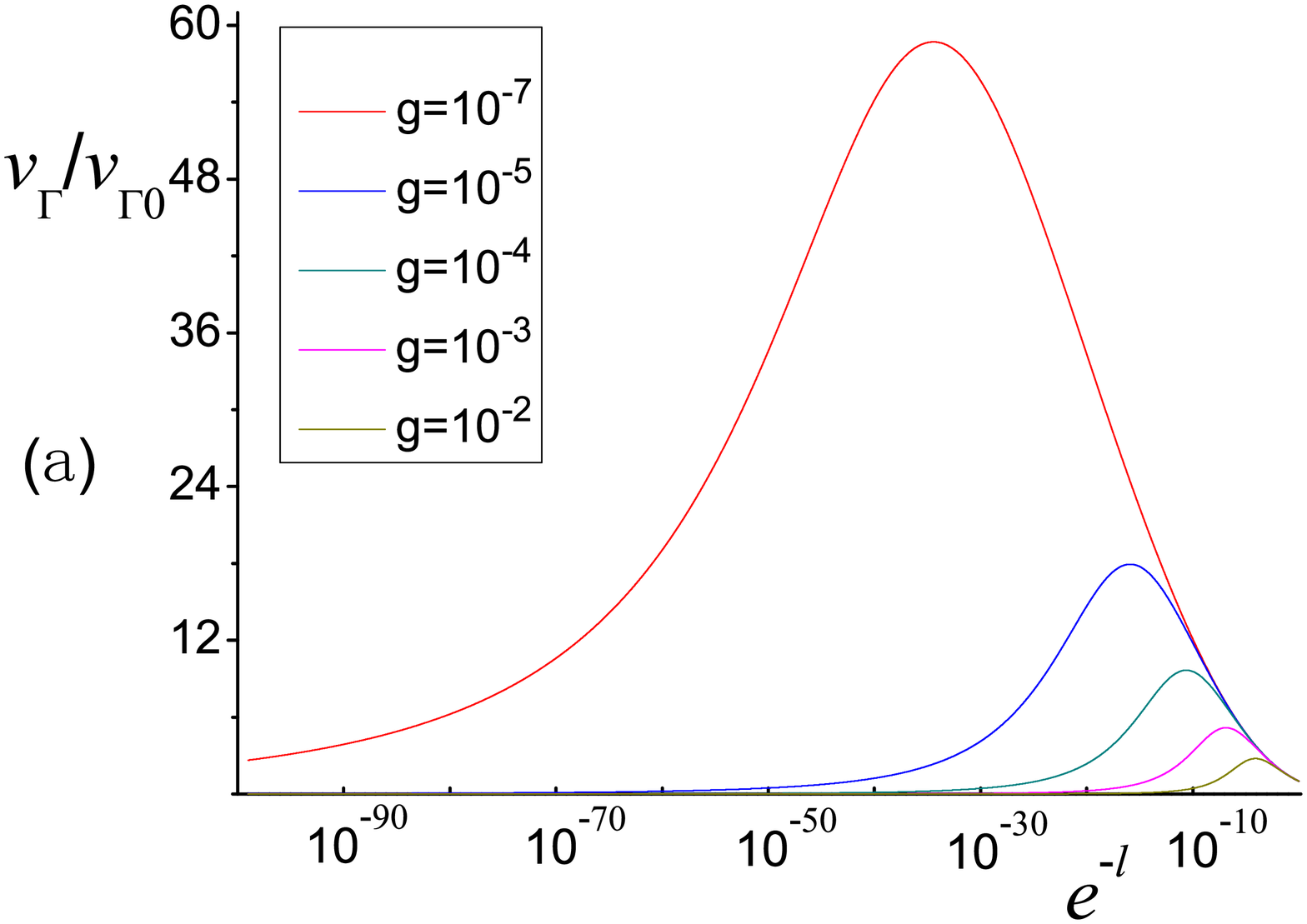,height = 5.6cm,width=7.6cm}
       ~\\
       \vspace{-0.85cm}
       \epsfig{file=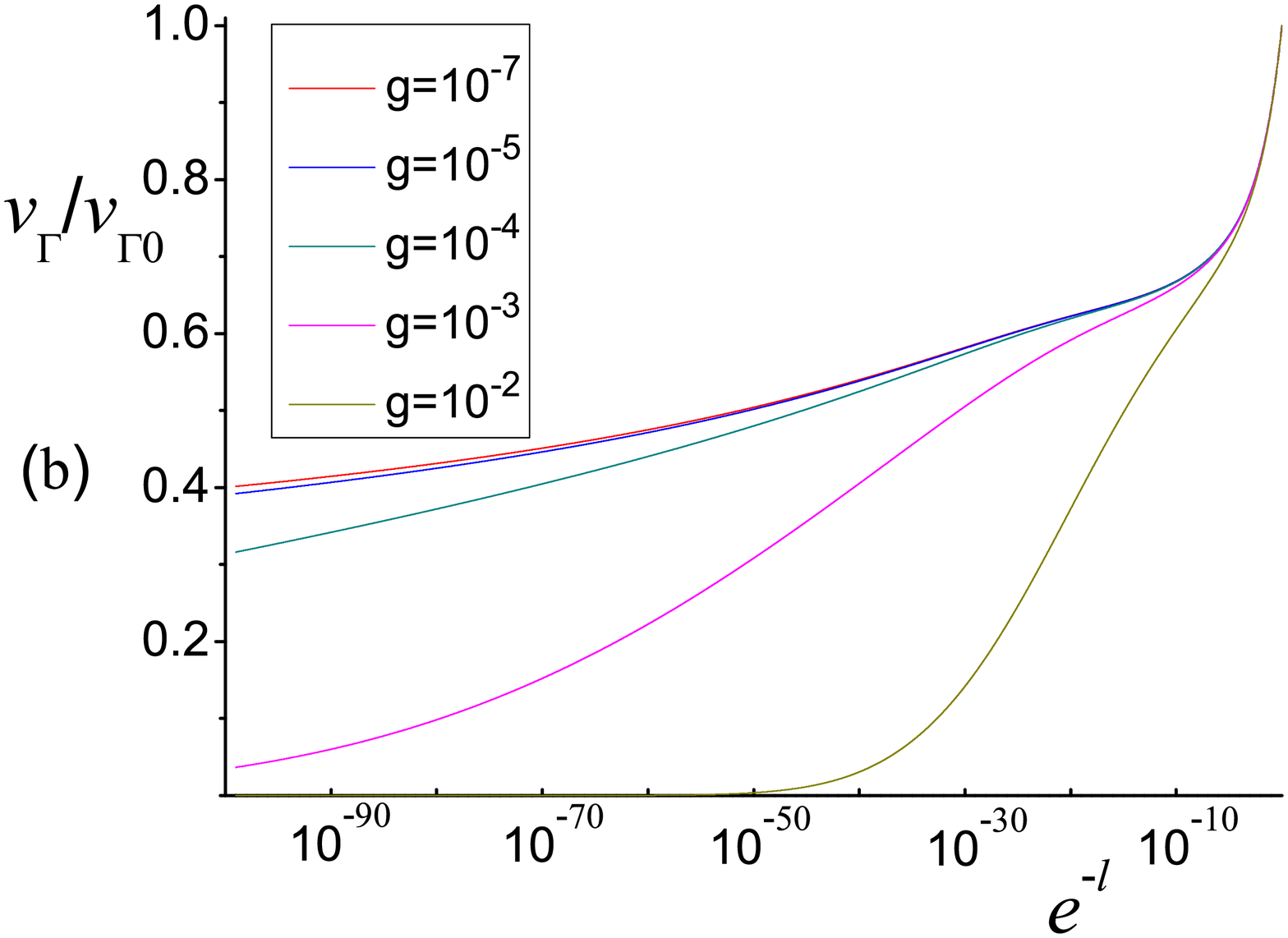,height = 5.6cm,width=7.6cm}
       ~\\
       \vspace{-0.80cm}
       \epsfig{file=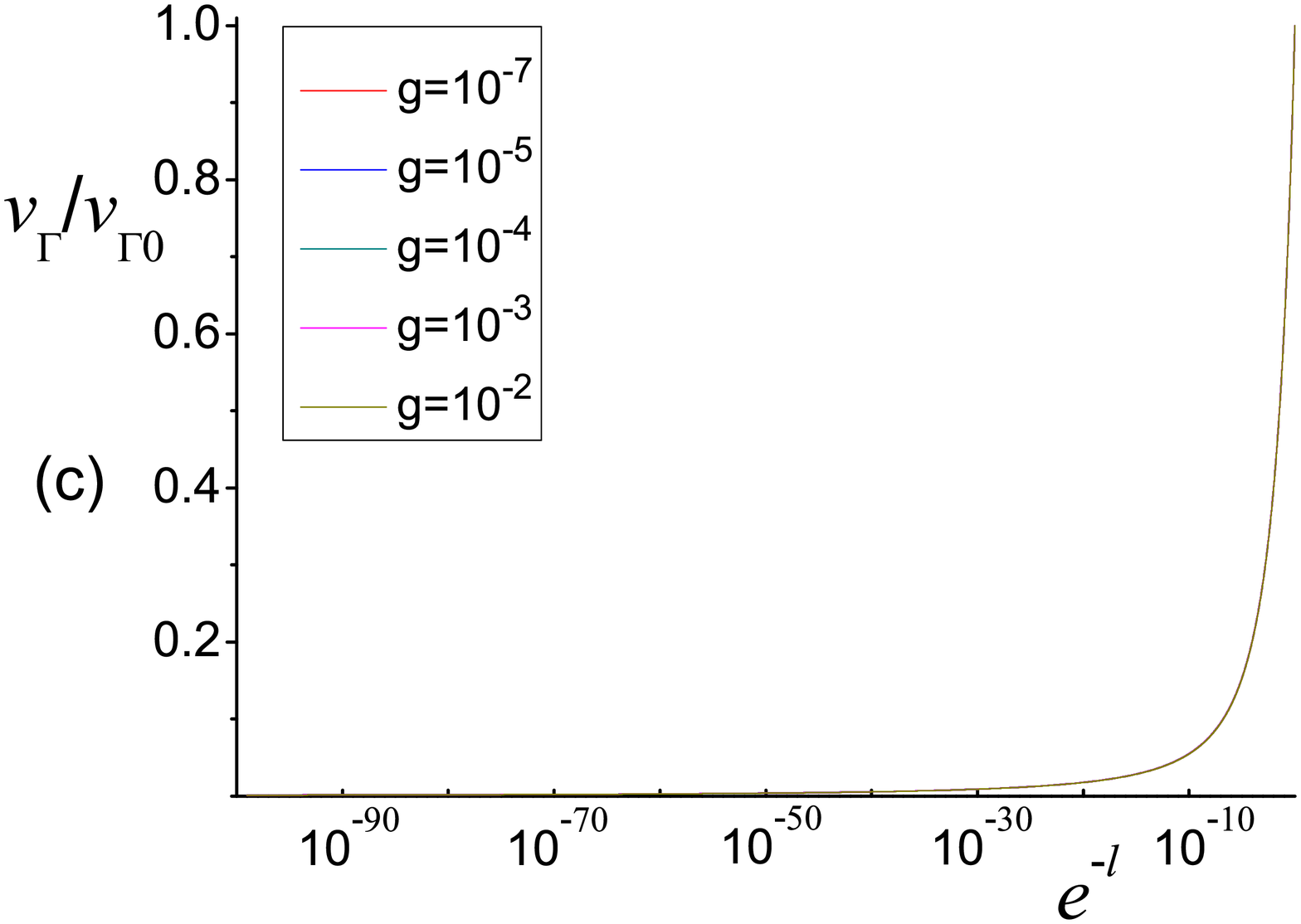,height = 5.6cm,width=7.6cm}
       ~\\
\vspace{-0.60cm} \caption{(a) $v_\Gamma$ for random mass; (b)
$v_\Gamma$ for random gauge potential $\tau^x$ component; (c)
$v_\Gamma$ for random gauge potential $\tau^z$
component.}\label{v_Gamma}
\end{figure}

\section{Stability of extreme anisotropy against disorders}\label{Stability_discuss}

The RG flows of fermion velocities $v_F$ and $v_\Delta$ with growing
scale $l$ can be obtained by numerically solving the corresponding
coupled equations with the initial values $v_{F0}$, $v_{\Delta0}$,
and $ v_{\Gamma0}$. First of all, in the clean limit $g=0$, the
equations reduce to that obtained by Huh and Sachdev. In this case,
it was already known that an extreme anisotropy of fermion
velocities, i.e., $v_\Delta/v_F \rightarrow 0$, is caused by nematic
order parameter fluctuation. The effects of various disorders on
this fixed point will be transparent when the dimensionless
parameter $g$ is increased smoothly.

\begin{figure}[t]
  \centering
     \epsfig{file=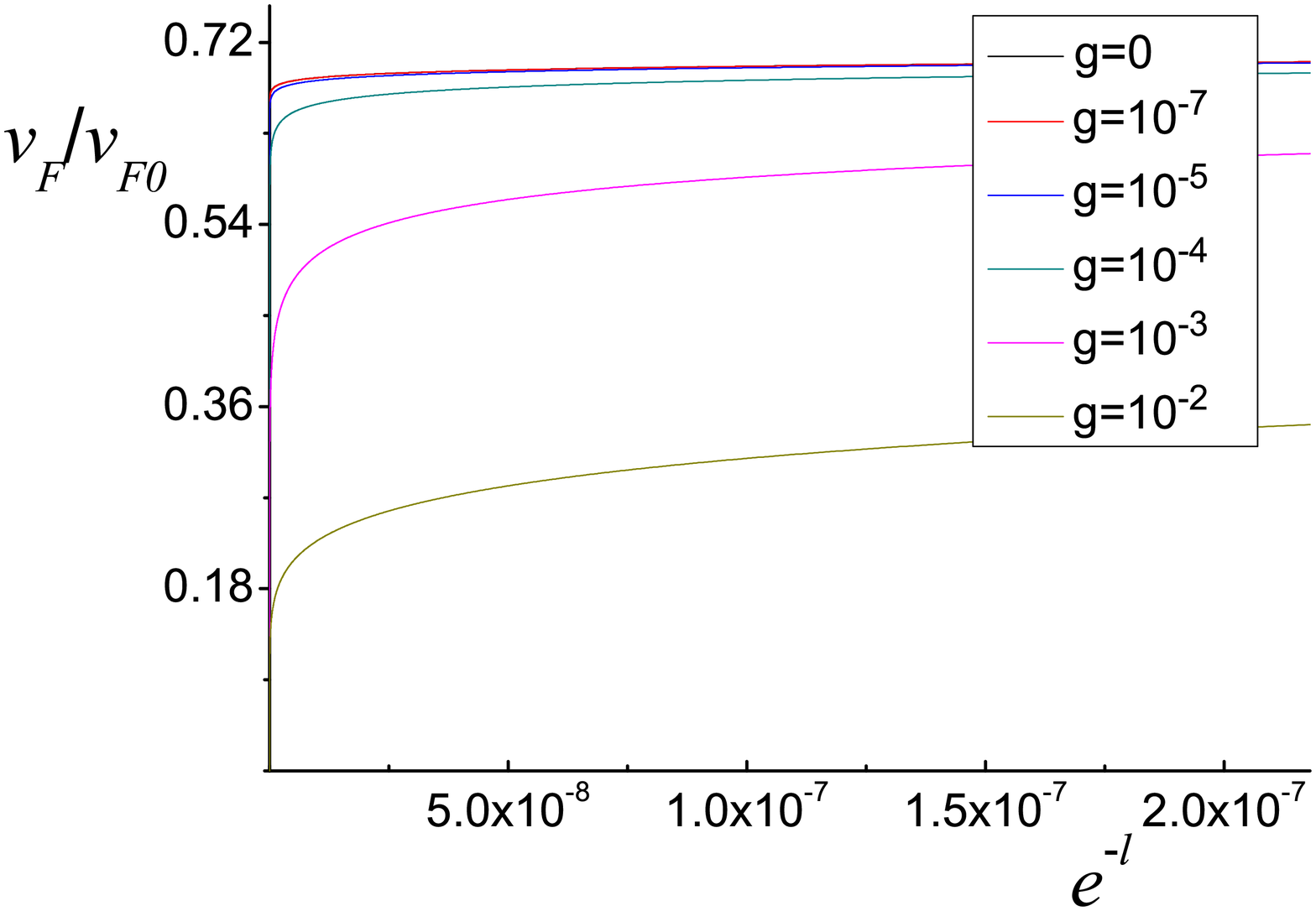,height=5.6cm,width=7.6cm}
     ~\\
     \vspace{-0.75cm}
     \epsfig{file=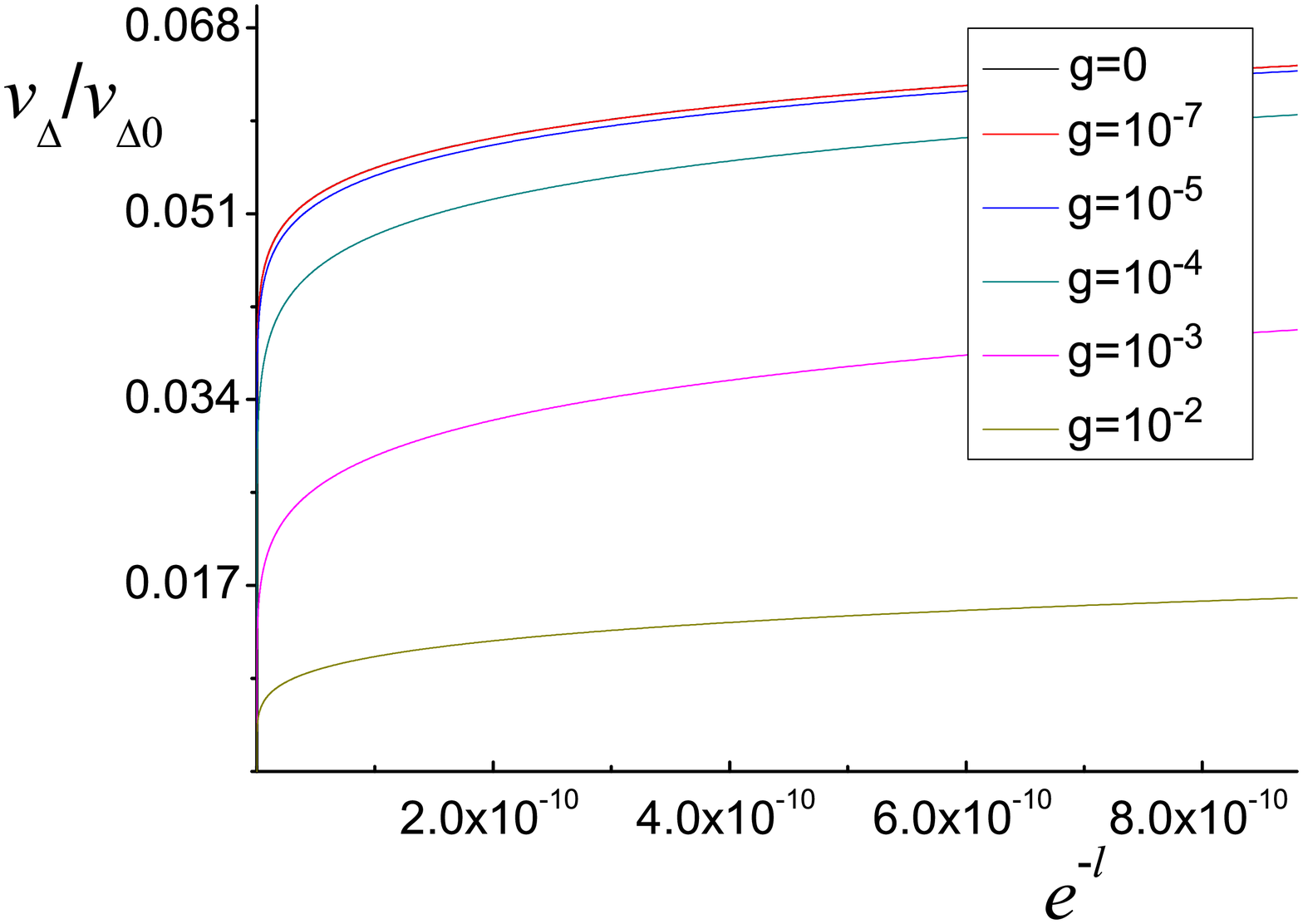,height=5.6cm,width=7.6cm}
     ~\\
     \vspace{-0.50cm}
    \caption{The flows of $v_F$ (upper one) and $v_\Delta$ (lower one) in the case of random
    mass.}
    \label{v_mass}
\end{figure}

We first consider the case of random mass. As $l$ grows, $v_\Gamma$
first increases and then decreases, eventually approaching zero, as
shown in Fig. \ref{v_Gamma}(a). Although the RG equations for
fermion velocities are modified by scattering due to random mass,
the disorder parameter $v_\Gamma$ flows to zero as $l \rightarrow
\infty$. Apparently, the random mass is irrelevant in the present
problem. As can be easily seen from see Fig. \ref{v_mass}, the
fermion velocities $v_F$ and $v_\Delta$ both decrease as $l$ grows.
More concretely, $v_\Delta$ goes down to zero rapidly, but $v_F$
decreases much more slowly and actually approaches a finite value.
These results imply the existence of extreme velocity anisotropy
with $v_\Delta/v_F \rightarrow 0$ in the presence of random mass.

\begin{figure}[t]
   \centering
       \epsfig{file=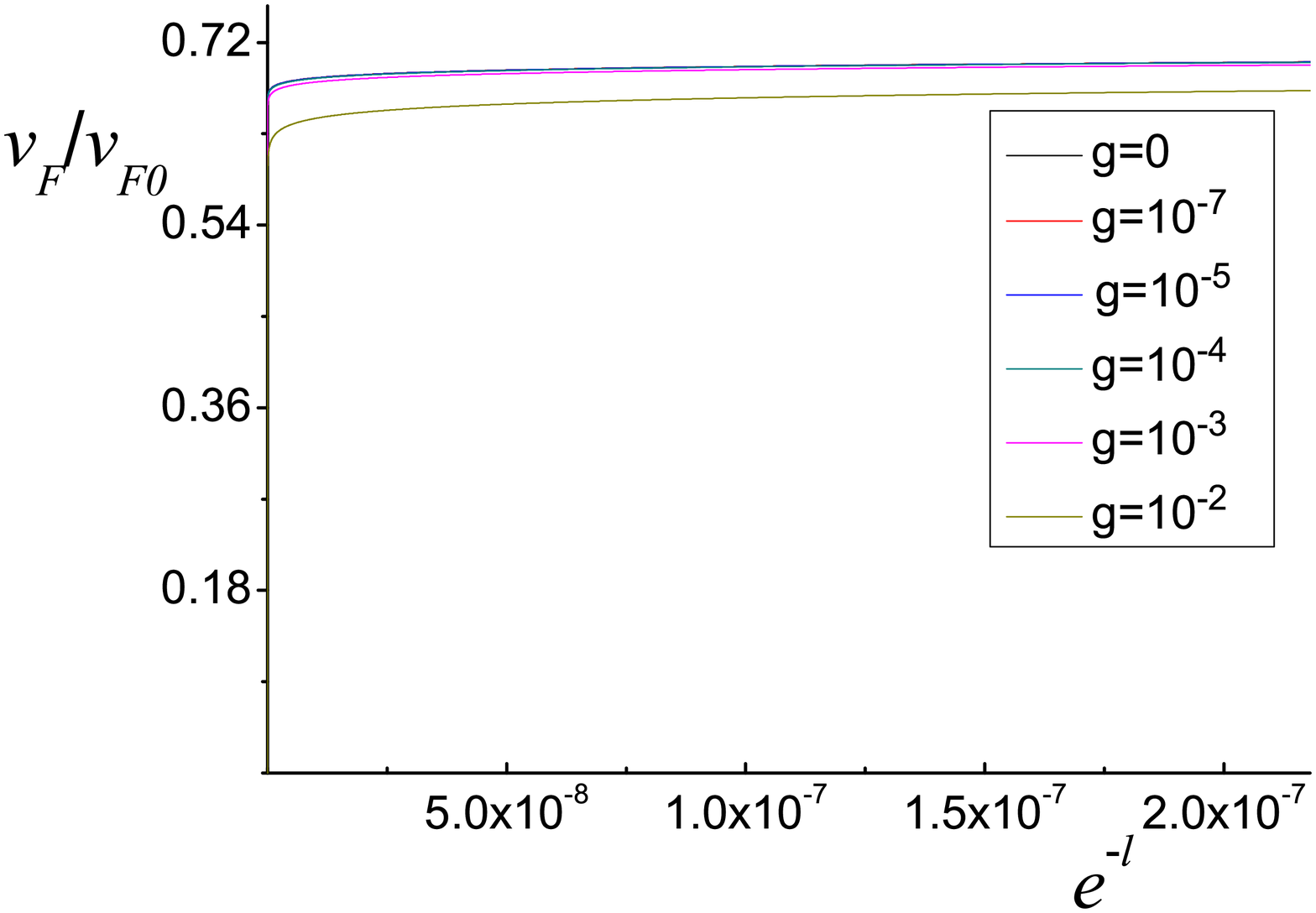,height=5.6cm,width=7.6cm}
       ~\\
       \vspace{-0.75cm}
       \epsfig{file=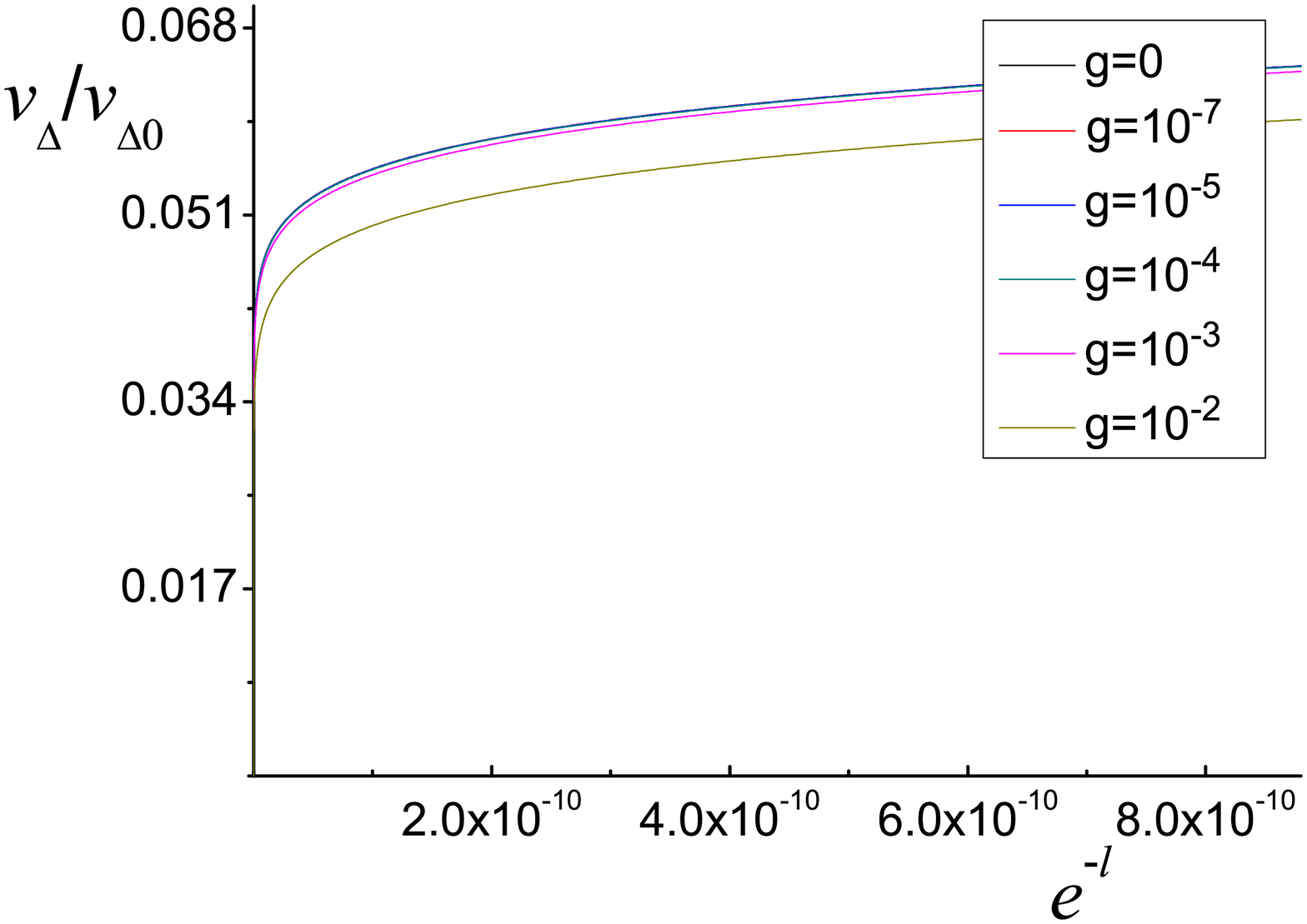,height=5.6cm,width=7.6cm}
       ~\\
       \vspace{-0.50cm}
 \caption{The running $v_F$ and $v_\Delta$ for the $\tau^x$ component of random gauge
 potential. The running of fermion velocities for component $\tau^z$ are very
 similar to this case, and thus are not shown.} \label{v_F_gauge}
\end{figure}

We next discuss the case of random gauge potential. The flows of
disorder strength $v_\Gamma$ with growing $l$ are shown in Fig.
\ref{v_Gamma}(b) for component $\tau^x$ and in Fig. \ref{v_Gamma}(c)
for component $\tau^z$. For both components, the corresponding
$v_\Gamma$ decrease as $l$ grows and finally approaches zero as $l
\rightarrow \infty$. Similar to the case of random mass, the random
gauge potential makes no important contributions to the flow of
fermion velocities. Therefore, as in the clean limit, both $v_F$ and
$v_\Delta$ decreases with $l$ until approaching zero for component
$\tau^x$ depicted in Fig. \ref{v_F_gauge}; the flows of $v_F$ and
$v_\Delta$ for component $\tau^z$ with growing $l$ will not be shown
since they are similar to those in the case of component $\tau^x$.
It is obvious that random gauge potential does not change the
extreme velocity anisotropy.

Unlike random mass and random gauge potential, the disorder strength
parameter $v_\Gamma$ in the case of random chemical potential does
not flow with $l$ and thus should be kept as a constant. As such,
the influence of scattering due to random chemical potential can not
be neglected and indeed the flows of velocities $v_F$ and $v_\Delta$
depend heavily on the magnitude of $v_\Gamma$. At first glance, the
flow equation of velocity ratio $v_\Delta/v_F$ is independent of
disorder strength $v_\Gamma$, as shown in Eq. (55), and thus appears
to have fixed point at $v_\Delta/v_F = 0$ as in the clean limit.
However, this solution is artificial. In the present problem, the
flow equation of $v_\Delta/v_F$ is derived from the more fundamental
equations of $v_\Delta$ and $v_F$, and therefore is reliable only
when $v_\Delta$ and $v_F$ have well-defined fixed points. If the
equations for $v_\Delta$ and $v_F$ have no fixed points, the
equation of $v_\Delta/v_F$ becomes meaningless.

To see the effect of random chemical potential, we make a
qualitative analysis based on the RG equations of fermion velocities
$v_\Delta$ and $v_F$. The fixed points can be obtained by requiring
that
\begin{eqnarray}
\frac{dv_F}{dl} &=& (C_1-C_2)v_F - \frac{v^2_\Gamma g}{2\pi
v_\Delta} = 0, \label{66} \\
\frac{dv_\Delta}{dl} &=& (C_1-C_3)v_\Delta - \frac{v^2_\Gamma
g}{2\pi v_F} = 0.\label{67}
\end{eqnarray}
We assume that $v_F^{*}$ and $v_\Delta^{*}$ correspond to the fixed
points. If both $v_F^{*}$ and $v_\Delta^{*}$ are finite, then the
above equations imply that $(C_1-C_2)v_\Delta^{*} v_F^{*} =
(C_1-C_3)v_\Delta^{*} v_F^{*}$, which can not be satisfied since
$v_\Delta^{*} \neq 0$. If $v_\Delta^{*} = 0$, then
\begin{eqnarray}
v_F^{*} = \frac{v^2_\Gamma g}{2\pi v_\Delta^{*}(C_1 - C_2)}.
\end{eqnarray}
From the expressions for $C_1$ and $C_2$, this implies that $1
\propto 1/(v_\Delta^{*})^2$, which is clearly inconsistent with the
assumption of $v_\Delta^{*} = 0$. Before going to $v_F^{*} = 0$
case, we define $D_i=(v_F/v_\Delta)C_i,i=(1,2,3)$, then the new
forms of Eq.(\ref{66}) and Eq.(\ref{67}) become
\begin{eqnarray}
(D_1-D_2)v_\Delta&=&\frac{v^2_\Gamma g}{2\pi v_\Delta} , \\
(D_1-D_3)\frac{v_\Delta^2}{v_F}&=&\frac{v^2_\Gamma g}{2\pi v_F} .
\end{eqnarray}
If $v_F^{*} = 0$, by both analytical and numerical analysis, we
found that these equations have no solution.

In summary, as shown by the above analysis, there is no fixed point
of the fermion velocities $v_F$ and $v_\Delta$ when the fermions
interact with random chemical potential. Therefore, $v_F$ and
$v_\Delta$ do not approach any stable values at the low energy
regime. Alternatively, straightforward numerical calculations show
that they oscillate rapidly between positive and unphysical negative
values as $l$ grows. In this case, the extreme velocity anisotropy
fixed point is destroyed. We interpret the occurrence of unphysical
negative velocities as a signature of the instability of nematic
phase transition in the presence of random chemical potential.

\section{Summary and Discussion}\label{summary}

In summary, we have examined the disorder effect near the critical
point of nematic phase transition in $d$-wave cuprate
superconductor. We considered three types of quenched disorders that
couple directly to the gapless nodal quasiparticles: random mass,
random gauge field, and random chemical potential. By means of a RG
analysis, we have derived a series of self-consistent flow equations
for fermion velocities and disorder strength. It was found that the
fixed point of extreme velocity anisotropy due to critical
fluctuation of nematic order parameter is not changed by random mass
and random gauge field, which are both irrelevant at low energy.
Therefore, it seems reasonable to expect an enhancement of dc
thermal conductivity at low temperature if there are only these two
kinds of disorders. However, when there is moderately strong random
chemical potential, which is marginal, the extreme anisotropy fixed
point is destroyed. Moreover, the nematic phase transition may
become unstable in the presence of such random chemical potential.

The extreme velocity anisotropy produced by the critical nematic
fluctuations may be probed by the heat transport measurements, since
it leads to a remarkable enhancement of the low-temperature thermal
conductivity. Apart from transport measurements, such extreme
anisotropy can also show its existence in angle-resolved
photo-emission spectroscopy (ARPES) experiments. Indeed, the
currently known value of the velocity ratio $v_\Delta/v_F$ in
$d$-wave cuprate superconductors was extracted from both heat
transport \cite{Chiao} and ARPES measurements \cite{Mesot}. Unlike
heat transport experiments that can only estimate the ratio
$v_\Delta/v_F$, the ARPES measurements are able to determine the
Fermi velocity $v_F$ and the gap velocity $v_\Delta$ separately
\cite{Mesot}. From the solutions of RG equations, we know that, the
extreme velocity anisotropy emerges because $v_\Delta$ is driven by
the critical nematic fluctuations to drop rapidly down to zero at
large $l$ but $v_F$ is driven to decrease very slowly. It should be
possible to detect the extreme anisotropy by measuring $v_F$ and
$v_\Delta$ separately by means of the ARPES experiments.

In addition, the effects of the nematic order parameter fluctuations
can also be reflected in the single-particle spectral function of
nodal quasiparticles. Kim $et$ $al.$ \cite{Kim} investigated this
issue and found two important features: strong angle-dependence of
quasipaticle scattering, and an enhancement of velocity anisotropy.
These predicted features of the fermion spectral function are
expected to be tested by ARPES experiments.

We next would like to remark on the disorder effects on the
polarization function and the final results. In our calculations,
the polarization function is obtained from the bubble diagram shown
in Fig. \ref{nem_polarization}, since including internal disorder
scattering line will introduce an additional suppressing factor $g
v_\Gamma^2$. To justify this approximation, we now make a
qualitative analysis on the disorder effects. It is well known that
disorder scattering generates a fermion damping rate $\gamma_0$,
which shifts the energy of Dirac fermions from $\omega$ to $\omega +
i\gamma_0$. In the presence of finite $\gamma_0$, it seems possible
to get a full analytical expression for the polarization function
$\Pi(q_x,q_y,\epsilon)$ only in the static ($\epsilon = 0$) limit.
Following the computational procedures given in \cite{Liu, Liu2}, we
have
\begin{eqnarray}
\Pi(q_x,q_y,\gamma_0) &=& \frac{1}{2\pi^2v_Fv_\Delta}\int^1_0dx
\frac{2\sqrt{x(1-x)}v_F^2q_x^2}{\sqrt{v_F^2q_x^2+v_\Delta^2q_y^2}}
\nonumber \\
&& \times \arctan \left(\gamma_0^{-1}\sqrt{x(1-x)(v_F^2
q_x^2+v_\Delta^2q_y^2)}\right) \nonumber \\
&& + (q_x \longleftrightarrow q_y).
\end{eqnarray}
In order to simplify this expression, we now consider the low-energy
regime, $|\mathbf{q}|\leq \gamma_0$, which leads to $\arctan
\left(\gamma_0^{-1} \sqrt{x(1-x)(v_F^2 q_x^2 +
v_\Delta^2q_y^2)}\right) \approx \gamma_0^{-1} \sqrt{x(1-x)(v_F^2
q_x^2 + v_\Delta^2q_y^2)}$. Using this approximation, the
polarization function is found to be
\begin{eqnarray}
\Pi(q_x,q_y,\gamma_0) = \frac{v_F}{6\pi^2v_\Delta
\gamma_0}(q_x^2+q_y^2).
\end{eqnarray}
Substituting this new polarization to the nematic propagator and
then performing the same RG calculations, we found that the
qualitative results presented in Sec. \ref{Stability_discuss} do not
change. We thus conclude that it is justified to neglect disorder
effects in the polarization function for weak disorders. Admittedly,
when disorders are strong enough to cause Anderson localization, the
RG approach utilized in our manuscript is no longer applicable and a
new RG scheme is needed to deal with the vertex corrections
generated by disorder scattering.

In this paper, we have considered only the coupling between
disorders and fermionic nodal quasiparticles since we are mainly
interested in the disorder effects on the RG flow of fermion
velocities. In practice, the effects of disorders on the nematic
transition are more complicated. For instance, there might be
quenched disorders that couple directly to the nematic order
parameter. This issue was briefly discussed by Kim $et$ $al.$, who
argued \cite{Kim} that quenched disorder may smear the
symmetry-breaking type quantum phase transition thereby producing a
glassy state. It is currently unclear how the nodal quasiparticles
are influenced by such kind of disorders.

\section{Acknowledgments}

G.Z.L. and H.K. would like to thank Flavio S. Nogueira for valuable
discussions. J.W. is grateful to Wei Li and Jing-Rong Wang for their
helps. G.Z.L. acknowledges the financial support from the National
Science Foundation of China under Grant No.11074234 and the project
sponsored by the Overseas Academic Training Funds of University of
Science and Technology of China.

\section*{Appendix}

In order to maintain a self-consistency of this paper, here we
provide a detailed calculation of the fermion self-energy (16).
Following Ref. \cite{Huh}, we can define
\begin{widetext}
\begin{eqnarray}
\Sigma_{\mathrm{nm}}(K) &=&
\int\frac{d^3Q}{(2\pi)^3}F(Q+K)G(Q)\mathcal
{K}\left(\frac{(\mathbf{q+k})^2}{\Lambda^2}\right)\mathcal
{K}\left(\frac{\mathbf{q}^2}{\Lambda^2}\right),
\end{eqnarray}
where $K\equiv (\mathbf{k},\omega)$ and $Q\equiv(\mathbf{q},\epsilon)$ are 3-momenta.
Here $\mathcal {K}(y)$ is an arbitrary function with $\mathcal
{K}(0)=1$, and it falls off rapidly with $y$, e.g. $\mathcal
{K}(y)=e^{-y}$. However, the results are independent of the
particular choices of $\mathcal {K}(y)$. It is easy to identify
that,
\begin{eqnarray}
G(Q) &=& \frac{1}{\Pi(\mathbf{q},\epsilon)}, \\
F(Q+K) &=& \frac{1}{N_f}\frac{i(\epsilon+\omega)-v_F(q_x+k_x)\tau^z
+ v_\Delta (q_y+k_y)\tau^x}{(\epsilon + \omega)^2+v_F^2(q_x+k_x)^2 +
v_\Delta^2 (q_y+k_y)^2}.
\end{eqnarray}
Expanding $F(Q+K) \mathcal{K}
\left(\frac{(\mathbf{q+k})^2}{\Lambda^2}\right)$ at $Q+K=Q$, and
retaining the first order, we have
\begin{eqnarray}
&&F(Q+K)\mathcal {K}\left(\frac{(\mathbf{q+k})^2}{\Lambda^2}\right) \nonumber \\
&\approx& \left. K_\mu \left[ \frac{\partial F(Q+K)}{\partial Q_\mu}
\mathcal {K}\left(\frac{(\mathbf{q+k})^2}{\Lambda^2}\right) +
F(Q+K)\frac{2(\mathbf{q+k})_\mu}{\Lambda^2}\mathcal
{K'}\left(\frac{(\mathbf{q+k})^2}{\Lambda^2}\right)\right]\right|_{Q_\mu+K_\mu=Q_\mu}
\nonumber\\
&=& K_\mu\left[\frac{\partial F(Q)}{\partial Q_\mu}\mathcal
{K}\left(\frac{\mathbf{q}^2}{\Lambda^2}\right)+F(Q)\frac{2q_\mu}{\Lambda^2}\mathcal
{K'}\left(\frac{\mathbf{q}^2}{\Lambda^2}\right)\right].
\end{eqnarray}
Then the self-energy becomes
\begin{eqnarray}
\Sigma_{\mathrm{nm}}(K)
&\approx&K_\mu\int\frac{d^3Q}{(2\pi)^3}\left[\frac{\partial
F(Q)}{\partial Q_\mu}G(Q)\mathcal
{K}^2\left(\frac{\mathbf{q}^2}{\Lambda^2}\right)\right. +
\left.F(Q)G(Q)\frac{2q_\mu}{\Lambda^2}\mathcal
{K}\left(\frac{\mathbf{q}^2}{\Lambda^2}\right)\mathcal
{K'}\left(\frac{\mathbf{q}^2}{\Lambda^2}\right) \right],
\end{eqnarray}
which leads to
\begin{eqnarray}
\frac{d \Sigma_{\mathrm{nm}}(K)}{d \Lambda} &=&
K_\mu\int\frac{d^3Q}{(2\pi)^3}\left\{\Bigl[\frac{-4\mathbf{q}^2}{\Lambda^2}\frac{\partial
F(Q)}{\partial Q_\mu}-4F(Q)\frac{q_\mu}{\Lambda^2}
\Bigl]G(Q)\mathcal {K}\Bigl(\frac{\mathbf{q}^2}{\Lambda^2}\Bigl)\mathcal
{K'}\Bigl(\frac{\mathbf{q}^2}{\Lambda^2}\Bigl)\right.
\nonumber\\
&&
-\left.\frac{4\mathbf{q}^2q_\mu}{\Lambda^4}F(Q)G(Q)\Bigl[\mathcal
{K}\Bigl(\frac{\mathbf{q}^2}{\Lambda^2}\Bigl)\mathcal
{K''}\Bigl(\frac{\mathbf{q}^2}{\Lambda^2}\Bigl)
+\mathcal {K'}^2\Bigl(\frac{\mathbf{q}^2}{\Lambda^2}\Bigl)\Bigl]\right\}.
\end{eqnarray}
It is convenient to introduce the following cylindrical coordinates,
\begin{eqnarray}
Q_\mu &=& y\Lambda(v_Fx,\cos\theta,\sin\theta),\nonumber\\
\hat{Q}_\mu &=& (v_Fx,\cos\theta,\sin\theta),\nonumber\\
q_\mu &=& y\Lambda(0,\cos\theta,\sin\theta),\nonumber\\
\hat{q}_\mu &=& (0,\cos\theta,\sin\theta),\nonumber\\
d^3Q &=& y^2\Lambda^3v_F dxdyd\theta.\nonumber
\end{eqnarray}
It is straightforward to obtain
\begin{eqnarray}
F(\hat{Q}) &=& \frac{1}{N_fv_F}\left(\frac{ix-\cos\theta\tau^z +
(v_\Delta/v_F)\sin\theta\tau^x}{x^2+\cos^2\theta+(v_\Delta/v_F)^2
\sin^2\theta}\right), \nonumber \\
G(\hat{Q}) &=& \frac{1}{\Pi(\hat{Q})} = 16v_\Delta
\mathcal{G}(x,\theta),
\end{eqnarray}
where
\begin{eqnarray}
\mathcal{G}^{-1} &=& \frac{x^2+\cos^2\theta}{\sqrt
{x^2+\cos^2\theta+(v_\Delta/v_F)^2 \sin^2\theta}} \nonumber \\
&& + \frac{x^2+\sin^2\theta}{\sqrt{x^2 +
\sin^2\theta+(v_\Delta/v_F)^2\cos^2\theta}}.
\end{eqnarray}
Since $F$ and $G$ are homogenous functions,
\begin{eqnarray}
F(Q)=\frac{1}{y\Lambda} F(\hat{Q}),\,\,\,\,\,G(Q)=\frac{1}{y\Lambda}
G(\hat{Q}).
\end{eqnarray}
We thus have
\begin{eqnarray}
\Lambda\frac{d \Sigma_{\mathrm{nm}}(K)}
{d \Lambda}&\approx&\frac{v_F K_\mu}{(2\pi)^3}\int^{\infty}_{-\infty}dx \int^{2\pi}_{0} d \theta \int^{\infty}_{0}dy\Bigl\{\Bigl[-4y
\frac{\partial F(\hat{Q})}{\partial \hat{Q}_\mu}-4y\hat{q}_\mu F(\hat{Q}) \Bigl]G(\hat{Q})\mathcal {K}(y^2)\mathcal {K'}(y^2)\nonumber\\
&&-4y^3\hat{q}_\mu F(\hat{Q}) G(\hat{Q})\Bigl[\mathcal {K}(y^2)\mathcal {K''}(y^2)
+\mathcal {K'}^2(y^2)\Bigl]\Bigl\}\nonumber\\
&=& \frac{v_F K_\mu}{8\pi^2}\int^{\infty}_{-\infty}dx
\int^{2\pi}_{0} d \theta \Bigl\{\Bigl[-4 \frac{\partial
F(\hat{Q})}{\partial \hat{Q}_\mu}-4\hat{q}_\mu F(\hat{Q})
\Bigl]G(\hat{Q})\int^{\infty}_0ydy\mathcal {K}(y^2)\mathcal
{K'}(y^2) \nonumber\\
&& -4\hat{q}_\mu F(\hat{Q}) G(\hat{Q})\int^{\infty}_0 y^3 dy
\Bigl[\mathcal {K}(y^2)\mathcal {K''}(y^2) +
\mathcal{K'}^2(y^2)\Bigl]\Bigl\}.
\end{eqnarray}
After integrating $y$ out, we find
\begin{eqnarray}
&&\int^{\infty}_0 y^3 dy \Bigl[\mathcal{K}(y^2)\mathcal{K''}(y^2)
+\mathcal {K'}^2(y^2)\Bigl] = -\int^{\infty}_0 y dy
\mathcal{K}(y^2)\mathcal {K'}(y^2) = \frac{1}{4}.
\end{eqnarray}
Therefore,
\begin{eqnarray}
\Lambda \frac{d \Sigma_{\mathrm{nm}}(K)}{d \Lambda}
&=&\frac{v_F K_\mu}{8\pi^2}\int^{\infty}_{-\infty}dx \int^{2\pi}_{0} d \theta
\Bigl\{\Bigl[\frac{\partial F(\hat{Q})}{\partial \hat{Q}_\mu}
+\hat{q}_\mu F(\hat{Q})\Bigl]G(\hat{Q})
- \hat{q}_\mu F(\hat{Q}) G(\hat{Q})\Bigl\}\nonumber\\
&=&\frac{v_F K_\mu}{8\pi^2}\int^{\infty}_{-\infty}dx \int^{2\pi}_{0}
d\theta\frac{\partial F(\hat{Q})}{\partial \hat{Q}_\mu}G(\hat{Q}).
\end{eqnarray}
Formally, the fermion self-energy function can be expanded as
\begin{eqnarray}
\frac{d \Sigma_{\mathrm{nm}}(K)}{d \ln\Lambda} =
C_1(-i\omega)+C_2v_F k_x \tau^z+C_3 v_\Delta k_y\tau^x.
\end{eqnarray}
When $K_0=\omega$ and $\hat{Q}_0=v_Fx$, we finally have
\begin{eqnarray}
C_1(-i\omega) &=& \frac{v_F\omega}{8\pi^3}\int^{\infty}_{-\infty}dx
\int^{2\pi}_{0} d \theta\frac{\partial F(\hat{Q})}{\partial v_F
x}G(\hat{Q}) \nonumber \\
&=& \frac{2v_\Delta\omega}{N_f \pi^3v_F}\int^{\infty}_{-\infty}dx
\int^{2\pi}_{0} d \theta \frac{i(x^2+\cos^2\theta+(v_\Delta/v_F)^2
\sin^2\theta)-2x(ix-\cos\theta\tau^z+(v_\Delta/v_F)
\sin\theta\tau^x) }{(x^2+\cos^2\theta+(v_\Delta/v_F)^2
\sin^2\theta)^2}\mathcal {G}(x,\theta) \nonumber\\
&=& \frac{2(v_\Delta/v_F)}{N_f \pi^3}\int^{\infty}_{-\infty}dx
\int^{2\pi}_{0}d \theta\frac{x^2-\cos^2\theta-(v_\Delta/v_F)^2
\sin^2\theta }{(x^2+\cos^2\theta+(v_\Delta/v_F)^2
\sin^2\theta)^2}\mathcal {G}(x,\theta)(-i\omega),
\end{eqnarray}
which directly leads to
\begin{eqnarray}
C_1 = \frac{2(v_\Delta/v_F)}{N_f \pi^3}\int^{\infty}_{-\infty} dx
\int^{2\pi}_{0} d \theta \frac{x^2-\cos^2\theta-(v_\Delta/v_F)^2
\sin^2\theta}{(x^2+\cos^2\theta+(v_\Delta/v_F)^2
\sin^2\theta)^2}\mathcal {G}(x,\theta).
\end{eqnarray}
$C_2$ and $C_3$ can be obtained similarly.
\end{widetext}

\end{document}